# Improving the spatial resolution of a BOTDA sensor using deconvolution algorithm

Li Shen, Zhiyong Zhao, Can Zhao, Hao Wu, Chao Lu, *Fellow, OSA*, and Ming Tang, *Senior Member, IEEE*

*Abstract*—Spatial resolution improvement from an acquired measurement using long pulse is developed for Brillouin optical time domain analysis (BOTDA) systems based on the total variation deconvolution algorithm. The frequency dependency of Brillouin gain temporal envelope is investigated by simulation, and its impact on the recovered results of deconvolution algorithm is thoroughly analyzed. To implement a reliable deconvolution process, differential pulse-width pair (DPP) technique is utilized to effectively eliminate the systematic BFS distortion stemming from the frequency dependency of temporal envelope. The width of the pulse pairs should be larger than 40 ns as is analyzed theoretically and verified experimentally. It has been demonstrated that the proposed method can realize flexible adjustment of spatial resolution with enhanced signal-to-noise ratio (SNR) from an established measurement with long pump pulse. In the experiment, the spatial resolution is increased to 0.5 m and 1 m with high measurement accuracy by using the deconvolution algorithm from the measurement of 60/40 ns DPP signals. Compared with the raw DPP results with the same spatial resolution, 9.2 dB and 8.4 dB SNR improvements are obtained for 0.5 m and 1 m spatial resolution respectively, thanks to the denoising capability of the total variation deconvolution algorithm. The impact of sampling rate on the recovery results is also studied. The proposed sensing system allows for distortion-free Brillouin distributed sensing with higher spatial resolution and enhanced SNR from the conventional DPP setup with long pulse pairs.

## I. Introduction

Brillouin optical time domain analysis (BOTDA) has been intensively studied in the past few decades [1]-[4], due to the ability of distributed strain and temperature measurement along the fiber. For a BOTDA system, spatial resolution is one of the most important parameters, which determines the minimum fiber length over which a perturbation can be distinguished. Usually, the spatial resolution can be improved with a shorter pump pulse width. However, the pump pulse width is practically limited to about 10 ns by the phonon relaxation time in optical fibers, corresponding to a 1-m spatial resolution. Pump pulse width below the phonon relaxation time will lead to weaker Brillouin scattering signal and broadened Brillouin gain spectrum (BGS) linewidth. As a result, the Brillouin frequency shift (BFS) determination accuracy declines dramatically.

Various approaches have been proposed to enhance the spatial resolution of BOTDA sensors. A widely used principle is the pre-excitation of acoustic wave in the sensing fiber, which is the basis for several techniques. The pulse pre-pump (PPP) scheme employs a pump pulse that consists of two parts [5], i.e. a long pedestal light for the excitation of acoustic wave, which is immediately followed by a narrow high-power pulse to sense the fiber. The width and power of the pedestal light should be optimized to excite the acoustic wave and avoid introducing excessive non-local distortion to the BGS. The continuous wave (CW) light is also used to excite the acoustic wave, and then followed by a short section of interrogation disturbance, such as a light intensity switch off in the dark pulse scheme [6], [7], or a π phase shift pulse in the Brillouin echoes system [8]. Restricted by the spontaneous Brillouin noise of the CW pre-excitation light, the sensing range is usually limited to a kilometer scale. In addition, the second echo introduced by the inertial feature of the acoustic wave will deteriorate the accuracy of BFS estimation. Another proposed method is the gain-profile tracing (GPT) technique [9], where a pump pulse fills the sensing fiber is abruptly terminated. The time derivative of the decreasing probe power at a given distance is proportional to the local Brillouin gain. For a GPT system, the spatial resolution is decided by the falling time of the pump pulse. It has been demonstrated that centimeter spatial resolution can be achieved with limited sensing range.

Another kind of commonly used method to achieve sub-meter spatial resolution over long distance is the differential pulse-width pair (DPP) technique [10]. In a DPP-BOTDA system, Brillouin gain signals are separately measured using two long pump pulses with a width difference, and the sensing signal is obtained by subtracting the two traces. In this way, the constraint of phonon relaxation time is avoided

This work was supported in part by the National Natural Science Foundation of China under Grants 61722108 and 61931010; in part by the National Key R&D Program of China 2018YFB1801002; and in part by Innovation Fund of WNLO. (*Corresponding author: Zhiyong Zhao*)

L. Shen, C. Zhao, H. Wu and M. Tang are with the Wuhan National Laboratory for Optoelectronics (WNLO) & National Engineering Laboratory for Next Generation Internet Access System, School of Optical and Electronic Information, Huazhong University of Science and Technology, Wuhan 430074, China (e-mail: shenli@hust.edu.cn; zhao_can@hust.edu.cn; wuhaoboom@qq.com; tangming@mail.hust.edu.cn).

Z. Zhao is with the School of Optical and Electronic Information, Huazhong University of Science and Technology, Wuhan 430074, China; He is also with the Photonics Research Centre, Department of Electronic and Information Engineering, The Hong Kong Polytechnic University, Hong Kong (zhiyong.zhao@polyu.edu.hk).

C. Lu is with the Photonics Research Centre, Department of Electronic and Information Engineering, The Hong Kong Polytechnic University, Hong Kong (e-mail: chao.lu@polyu.edu.hk).

by using long pump pulses, and the spatial resolution is determined by the pulse width difference. The DPP-BOTDA has the advantage of easy implementation, but the differential signal intensity attenuates with the improvement of spatial resolution.

The abovementioned methods need modifications of the system setup to change the spatial resolution. Alternatively, the signal post-processing methods, such as the deconvolution and subdivision algorithms [11]-[13], are proposed for flexible adjustment of the retrieved spatial resolution with single measurement. Both schemes treat the long pump pulse as a concatenation of several short pulses, and high spatial resolution can be realized by recovering the detailed sensing signal related to the short pulse. In reference [12], Richardson-Lucy iterative deconvolution algorithm is used to achieve a 5-m spatial resolution from a 100 ns pump pulse. Recently, the deconvolution algorithm has been used in the Brillouin distributed sensor [13], where 0.2-m spatial resolution can be retrieved with a 40 ns pump pulse, as well as a 1.7-fold improvement in the BFS accuracy over the DPP technique. However, it is found that BFS distortion will appear at fiber sections prior to the hotspot in the recovered results.

The BFS distortion is caused by the frequency dependency of Brillouin gain temporal envelope. Due to the fact that the distance of the distortion region from the hotspot is the same as the length determined by the pump pulse width, the distortion can be eliminated by combining the results of two long pump pulses with different durations. Although this method has been validated for single hotspot condition [13], it may not be applicable for multiple hotspots, where one hotspot may take place at the BFS distortion region of another hotspot. Moreover, the impact of temporal envelope on the deconvolution results of BOTDA signal has not been thoroughly analyzed yet.

In this work, to fundamentally eliminate the BFS distortion caused by the frequency dependency of temporal envelope, a modified deconvolution scheme based on the DPP pre-processing is presented. Thanks to the differential processing of DPP technique, the BFS distortion can be effectively avoided after deconvolution calculation. Theoretical and experimental results demonstrate that a BFS profile with enhanced spatial resolution can be recovered without distortion from 60/40 ns DPP signal. To further improve the reconstruction accuracy, total variation deconvolution algorithm is utilized, which allows for tunable adjustment of spatial resolution and measurement signal-to-noise ratio (SNR). In the experiment, 1-m and 0.5-m spatial resolution is retrieved by using the deconvolution algorithm. The obtained BFS profile is in great agreement with the conventional DPP result under identical spatial resolution. And a distinct SNR improvement is acquired with minor BFS degradation for the hotspots. Finally, the impact of sampling rate on the retrieved results of deconvolution algorithm is also investigated.

## II. Theoretical Model and Simulation Results

In a conventional BOTDA system, a pump pulse propagates along the fiber and interacts with the counter propagating CW probe light. Optical power will transfer between the pump pulse and the probe light if their frequency difference is near the local BFS. Usually, a rectangular pump pulse with width of $T_P$ is used. When the pump pulse passing through a short uniform section $\Delta z$ at fiber location $z$, the temporal response of induced Brillouin gain is expressed as $a_S^{short}(z,t)$ [14], [15]:

$$a_S^{short}(z,t) = p(z, t - z/V_g) h(z) \quad (1)$$

$$p(z,t) = \left\{1 - \exp\left[-\Gamma_A^*(z)t\right]\right\}\left[u(t) - u(t - T_P)\right] \quad (2)$$

$$h(z) = g(z)\frac{I_P^0 A_S^0}{2\Gamma_A^*(z)}\Delta z \quad (3)$$

$a_S^{short}(z,t)$ is the product of two terms, the first term $p(z,t-z/V_g)$ represents the temporal envelope of the induced Brillouin gain at position $z$, which has a width equal to the pump pulse and is dependent on the local frequency detuning parameter $\Gamma_A(z)$. The second term $h(z)$ is the impulse response of Brillouin gain. $g(z)$ is a constant related to the electrostrictive constant, $I_P^0$ is the pump pulse intensity, $A_S^0$ is the continuous constant of the probe light, $V_g$ is the light velocity in optical fibers, and $u(\cdot)$ represents the Heaviside unit step function. The frequency detuning parameter $\Gamma_A(z)$ can be expressed as [14], [15]:

$$\Gamma_A(z) = \frac{i\pi\left(v_B^2(z) - v^2 - iv\Delta v_B\right)}{v} \quad (4)$$

where $v_B(z)$ is the local BFS, $v$ is the frequency difference between the pump and probe light, and ($v$-$v_B$) is the detuned frequency. $\Delta v_B$ is the full width at half maximum (FWHM) of BGS, which is about 27 MHz for standard silica fibers.

### A. Characteristics of the Brillouin Gain Temporal Envelope

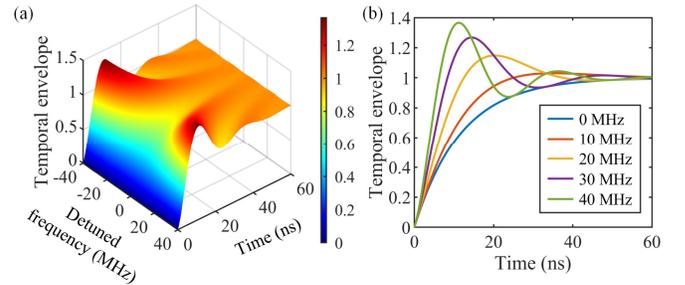

Fig. 1. (a) 3-D plot of the Brillouin gain temporal envelope at different detuned frequencies with a 60 ns pump pulse; (b) Exemplified temporal envelopes at some certain detuned frequencies.

According to Eq. 2, Fig. 1(a) illustrates the simulated 3-D plot of the real part of the Brillouin gain temporal envelope at different detuned frequencies with a 60 ns pump pulse. $v_B$ is 10.8 GHz, and $\Delta v_B$ is 27 MHz for the calculation. Exemplified temporal envelopes at some certain detuned frequencies are plotted in Fig. 1(b). When the detuned frequency is 0 MHz, which is at the peak frequency of BGS, the temporal envelope of Brillouin gain increases gradually by an exponential decay rate, while the temporal envelopes at other detuned frequencies shows quite different oscillations features. Noteworthy, the discrepancy of temporal envelope is mainly observed at the leading head of the pulses. The simulation result indicates that for pulse width exceeding 40 ns, the temporal envelope will reach a steady-state and converge to a constant for all the detuned frequencies. This conclusion also can be easily deduced from Eq. (2) with a large value of $t$.

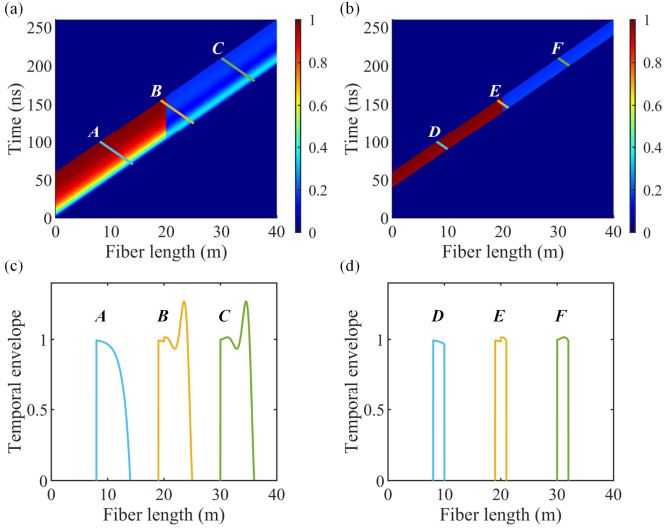

Fig. 2. (a) Temporal response of the induced Brillouin gain when a 60 ns pump pulse is launched; (b) Temporal response of the induced Brillouin gain for 60/40 ns DPP; (c) $p(z,t-2z/V_g)$ functions corresponding to the lines $A$, $B$, and $C$ of 60 ns pump pulse; (d) $p(z,t-2z/V_g)$ functions corresponding to the lines $D$, $E$, $F$ of 60/40 ns DPP.

Assume the sensing length is 40 m in the simulation, when the 60 ns pump pulse is launched into the fiber from $z=0$, the Brillouin gain will be generated along the fiber as the pump pulse propagates. Based on Eq. (1), Fig 2(a) shows the 2-D mapping of the induced Brillouin gain $a_S^{short}(z,t)$ at different positions along the fiber, which consists of two sections with different local BFSs. The BFS for the first 20 m is 10.8 GHz, which is the frequency difference between the pump and probe light in the simulation, and the BFS for the last 20 m is 10.83 GHz with a detuned frequency of 30 MHz. Due to the time delay of the backscattered light arriving at $z=0$, the received Brillouin gain signals are the integral of the fiber segments within the spatial resolution, as illustrated by the marked lines $A$, $B$ and $C$ in Fig. 2(a). So the received signal at time $t$ can be expressed as an integral along the lines:

$$r(t) = \int a_S^{short}(z, t-z/V_g)dz \\ = \int p(z, t-2z/V_g)h(z)dz \quad (5)$$

Eq. (5) shows that the received signal $r(t)$ is a weighted average of the impulse response $h(z)$. The weighting is determined by $p(z,t-2z/V_g)$ function, which is a combination of Brillouin gain temporal envelopes within the spatial resolution. Due to this integral process, the original sharp Brillouin gain change in $h(z)$ becomes a blurred transition in the received signal $r(t)$.

If temporal envelope $p(z,t)$ does not vary along the fiber, Eq. (5) can be regarded as a convolution process. However, $p(z,t)$ is dependent on the detuned frequency, thus it will spatially change with the local BFS. Figure 2(c) shows the $p(z,t-2z/V_g)$ functions corresponding to the lines $A$, $B$ and $C$ in Fig. 2(a). For a uniform fiber section with a constant BFS, such as the section covered by line $A$ and line $C$ in Fig. 2(a), $p(z,t-2z/V_g)$ is identical to the local temporal envelope. As for the BFS changing region, such as the one covered by line $B$ in Fig. 2(a), the corresponding $p(z,t-2z/V_g)$ becomes a concatenation of the temporal envelopes at different detuned frequencies. Since $p(z,t-2z/V_g)$ changes with the detuned frequency, if an invariant temporal envelope is used to implement deconvolution processing of Eq. (5), it will result in BFS distortion at some positions.

*B. Proposed Solution and Simulation Results*

Considering that the Brillouin gain temporal envelope at the part of pulse over 40 ns are nearly the same, the frequency dependent oscillation of the leading head can be cancelled by the subtraction of pulse pair with the width larger than 40 ns, yielding a time envelope independent of the detuned frequency. Different from the conventional DPP-BOTDA system to obtain a high spatial resolution, the ultra-short pulse width difference is not necessarily required in the proposed sensing system. So, in the simulation, 60/40 ns pulse pair is chosen to ensure high SNR of the received signal. The simulated temporal response of Brillouin gain of 60/40 ns DPP for the 40 m sensing fiber is shown in Fig. 2(b). Thanks to the DPP pre-processing, the frequency dependent fluctuation of Brillouin gain is removed. Figure 2(d) shows the $p(z,t-2z/V_g)$ function corresponding to the lines $D$, $E$ and $F$ in Fig. 2(b). Clearly, rectangle-like shapes can be kept for all positions. Therefore, it indicates that the deconvolution algorithm is promising to effectively recover the sensing signal without distortion, which is investigated in the simulations as follows.

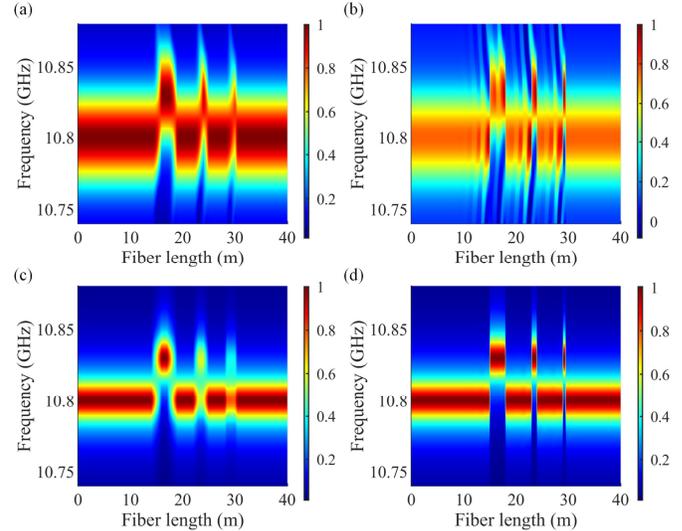

Fig. 3. (a) Simulated distribution of BGS with 20 ns pump pulse; (b) Normalized deconvolution result of 20 ns pump pulse; (c) Simulated distribution of BGS with 60/40 ns pump pulse pair; (d) Normalized deconvolution result of 60/40 ns pump pulse pair.

Assume there are three hotspots along the 40 m sensing fiber with length of 3 m, 1 m and 0.5 m, respectively, and the three hotspots undergo the same BFS change of 30 MHz with original BFS of 10.8 GHz. By modeling the Brillouin gain with Eq. (5), the received signal with different pump pulse schemes can be simulated. Figure 3(a) shows the distribution of BGS with a 20 ns pump pulse. The spatial resolution limitation can be observed as the 20 ns pump pulse is insufficient to identify the 1m and 0.5 m hotspots. Figure 3(b) shows the normalized direct deconvolution result with total variation deconvolution algorithm using the 20 ns pulse data. The temporal envelope at the peak frequency of BGS is used for the deconvolution process. Significant BGS errors can be clearly observed near the BFS changing regions.

As a comparison, the DPP signal is also processed by the deconvolution algorithm. The distribution of BGS of 60/40 ns DPP is shown in Fig. 3(c), which has a spatial resolution equal to the 20 ns pump pulse. The corresponding normalized deconvolution result based on the DPP signal is shown in Fig. 3(d). The temporal envelope of the DPP signal at the peak frequency of BGS is used for the recovery. As can be observed that, three hotspots with sharp edges can be recovered precisely without obvious errors.

To quantify the accuracy of the recovered results, the calculated BFS profiles of the 20 ns pump pulse and the corresponding results after deconvolution are shown in Fig. 4(a). The maximum systematic error before the hotspots is about 4.8 MHz, and the BFS of the hotspots cannot be accurately retrieved. Figure 4(b) shows the BFS profiles of the original 60/40 ns DPP signal and the corresponding results after deconvolution. The BFSs of the three hotspots are accurately recovered by deconvolution without any distortion. This simulation indicates that, deconvolution can be reliably implemented assisted with 60/40 ns DPP signal to improve the spatial resolution, avoiding the BFS distortion issue.

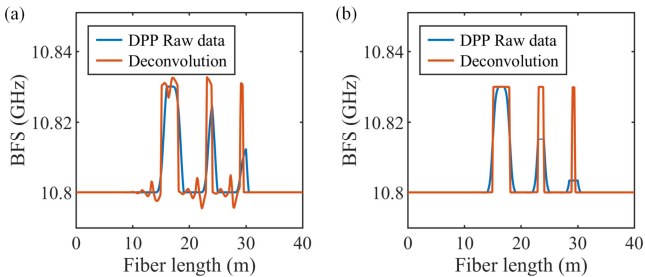

Fig. 4. (a) Calculated BFS profiles of the 20 ns pump pulse and the corresponding deconvoluted results; (b) Calculated BFS profiles of the 60/40 ns DPP signal and the corresponding deconvoluted results.

*C. Performance Analysis of the Total Variation Deconvolution Algorithm*

Above simulations are carried out without the consideration of noise. In practice, the received signal will be inevitably corrupted by noise. To reduce the influence of noise and obtain a more robust results, total variation deconvolution algorithm is explored in this work. For the BOTDA system, the received time domain signal $g$ can be modeled as [16]:

$$g = Hf + \eta \quad (6)$$

where $H$ is the convolution operator which is related to the $p(z, t-2z/V_g)$ function, $f$ is the unknown high spatial resolution signal, and $\eta$ is the additive noise. As has been simulated previously, for a conventional BOTDA signal, the convolution operator $H$ is dependent on the detuned frequency. For the DPP signal with an appropriate pulse width (i.e. > 40 ns), the convolution operator $H$ can be well represented by the temporal response at the peak frequency of BGS.

With a known $H$, simple inversion of Eq. (6) to solve $f$ is still ill-conditioned, because a small noise in the detected signal can cause a large perturbation in the restored results. This kind of problem can be alleviated by using regularization, which will however introduce additional restrictions to obtain a solution with acceptable precision. Total variation regularization solves the following minimization problem [16], [17]:

$$f = \arg\min \|Hf - g\|_2^2 + \mu \|f\|_{TV} \quad (7)$$

where $\mu > 0$ is the regularization parameter, and $\|\cdot\|_2$ is the $L_2$ norm. The function in Eq. (7) is a weighted sum of two terms. The fidelity term $\|Hf - g\|_2^2$ represents the difference with the received raw data, and more details can be preserved by reducing this term to obtain a higher spatial resolution. The regularization term $\|f\|_{TV}$ can be expressed as [16]:

$$\|f\|_{TV} = \sum_i |f_{i+1} - f_i| \quad (8)$$

$\|f\|_{TV}$ calculates the sum of variations between neighboring recovered data along the fiber length direction. The total variation deconvolution process is to find a solution of $f$ to minimize Eq. (7). The regularization parameter $\mu$ plays a critical role in determining the performance of deconvolution. When $\mu = 0$, the highest spatial resolution can be achieved without the restriction of regularization term, but the results are sensitive to noise. With the increasing of regularization parameter $\mu$, $\|f\|_{TV}$ plays an increasingly important role, which forces the result to be smooth. However, higher regularization parameter $\mu$ may over smooth the signal and remove some useful sensing information. Since the regularization parameter $\mu$ can be used to adjust the spatial resolution and the noise reduction effect, it should be carefully chosen according to the application requirement.

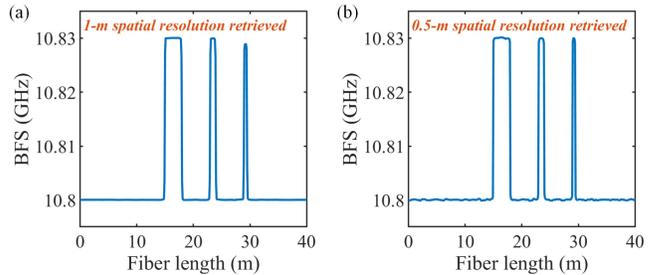

Fig. 5. (a) 1-m spatial resolutions retrieved from the 60/40 ns DPP signal; (b) 0.5-m spatial resolutions retrieved from the 60/40 ns DPP signal.

To illustrate the influence of the regularization parameter $\mu$, 1-m and 0.5-m spatial resolution results are retrieved from the 60/40 ns DPP signal used in the above simulations with used $\mu$ of 350 and 2200 respectively. The corresponding BFS profiles are shown in Fig. 5(a) and 5(b) respectively. The simulated raw data has a sampling rate of 1 GSa/s and SNR of 23 dB. Comparing with the 1-m retrieval results, when the retrieved spatial resolution is improved to 0.5 m, the BFS of the shortest hotspot is fully restored, but the BFS uncertainty along the fiber increases at the same time as shown by the fluctuations along the fiber. This result demonstrates that by adjusting the regularization parameter $\mu$, the spatial resolution can be improved at the expense of decreasing the measurement accuracy.

Comparing with the conventional deconvolution algorithm, the total variation deconvolution algorithm has the advantage of noise reduction. Therefore, the noise reduction performance of total variation deconvolution is also studied in this work. The side effect of noise reduction is that some useful sensing information may also be removed, which will result in a BFS degradation at hotspots. This phenomenon is more evident when higher SNR improvement is achieved. In the simulation, SNR improvement at two typical BFS degradation levels of 0.1 MHz and 0.5 MHz is investigated, with a sampling rate of 1 GSa/s. the raw data is the 60/40 ns DPP signal with original SNR of 23 dB. A hotspot along the fiber has a length varying from 0.5 m to 1.5 m. The regularization parameter $\mu$ is adjusted accordingly to recover the BFS of the hotspot with the target degradation level, then the length of hotspot is regard as the system spatial resolution. For each set of signals, deconvolution process is performed 100 times with randomly generated Gaussian white noise. The retrieved averaged BFS profile is used for degradation calculation to avoid the influence of noise. The SNR of the recovered results with different spatial resolution are shown in Fig. 6(a), which is calculated by the time-domain trace at the local peak Brillouin gain frequency. Meanwhile the simulated result of the conventional DPP technique is also presented for comparison. The result indicates that with acceptable BFS degradation level, the total variation deconvolution algorithm can effectively improve the SNR compared with the DPP results. With a BFS degradation tolerance of 0.1 MHz, a SNR improvement of 0.78-8.61 dB can be obtained with the spatial resolution from 0.5-1.5 m, while higher SNR improvement ranging from 7.70-16.39 dB can be achieved with a BFS degradation tolerance of 0.5 MHz.

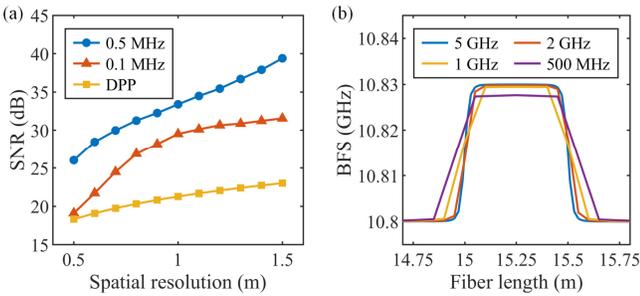

Fig. 6. (a) SNR of deconvolution results and DPP results at different spatial resolution; (b) Averaged BFS profile for a retrieved 0.5-m hotspot with different sampling rates.

Sampling rate is an important parameter that affects the recovered results. Figure 6(b) shows the recovered averaged BFS profile for a 0.5-m hotspot with different sampling rates. The raw data is a 60/40 ns DPP signal with 23 dB SNR. For different sampling rate, regularization parameter $\mu$ is chosen to obtain identical SNR of 26 dB after deconvolution. As can be seen, the BFS degradation of the hotspot decreases with the increased sampling rate. The BFS degradations are 2.34 MHz, 0.56 MHz, 0.22 MHz and 0.03 MHz respectively with sampling rate of 500 MSa/s, 1 GSa/s, 2 GSa/s and 5 GSa/s. The simulation shows that higher sampling rate can facilitate more accurate BFS recovery.

III. EXPERIMENTAL SETUP AND RESULTS

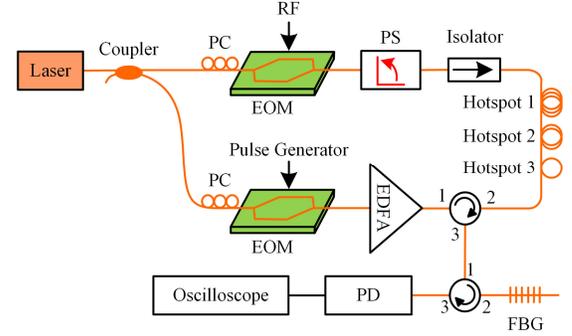

Fig. 7. Experimental setup of the BOTDA system. PC: polarization controller, EOM: electro-optic modulator, RF: radio frequency, PS: polarization switch, EDFA: erbium-doped fiber amplifier, FBG: fiber Bragg grating, PD: photodetector.

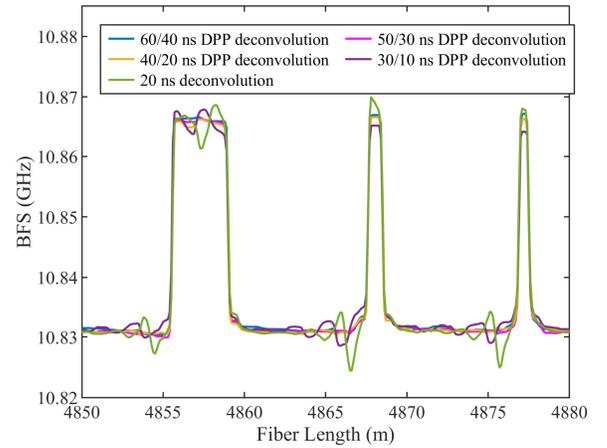

Fig. 8. Deconvolution results with different pump pulse schemes.

A typical BOTDA system is implemented to characterize the proposed deconvolution scheme, as illustrated in Fig. 7. A distributed feedback laser operating at 1550 nm is used to generate the probe and pump light through a 50:50 optical coupler. The probe light is modulated by an electro-optical modulator (EOM) to sweep the frequency, which is driven by a radio frequency (RF) generator through carrier suppressed double-sideband modulation. A polarization switch (PS) has been employed to reduce the polarization fading of Brillouin gain. Then the probe light is launched into a 4.9 km long sensing fiber after an isolator. Three hotspots at the fiber end with lengths of about 3.3 m, 1 m and 0.5 m are used to verify the spatial resolution of the system. At the lower branch, the light is modulated by another EOM to generate high extinction ratio pump pulse with fast rising/falling time by using a programmable electrical pulse generator. Amplified by an erbium-doped fiber amplifier (EDFA), the pump pulse is injected into the sensing fiber in a counter-propagating direction with respect to the probe light through an optical circulator. At the receiver side, a fiber Bragg grating (FBG) is employed to reflect the Brillouin Stokes sideband. Then the Brillouin signal is detected by a photodetector (PD), which is connected to a computer-controlled oscilloscope for automatic data acquisition.

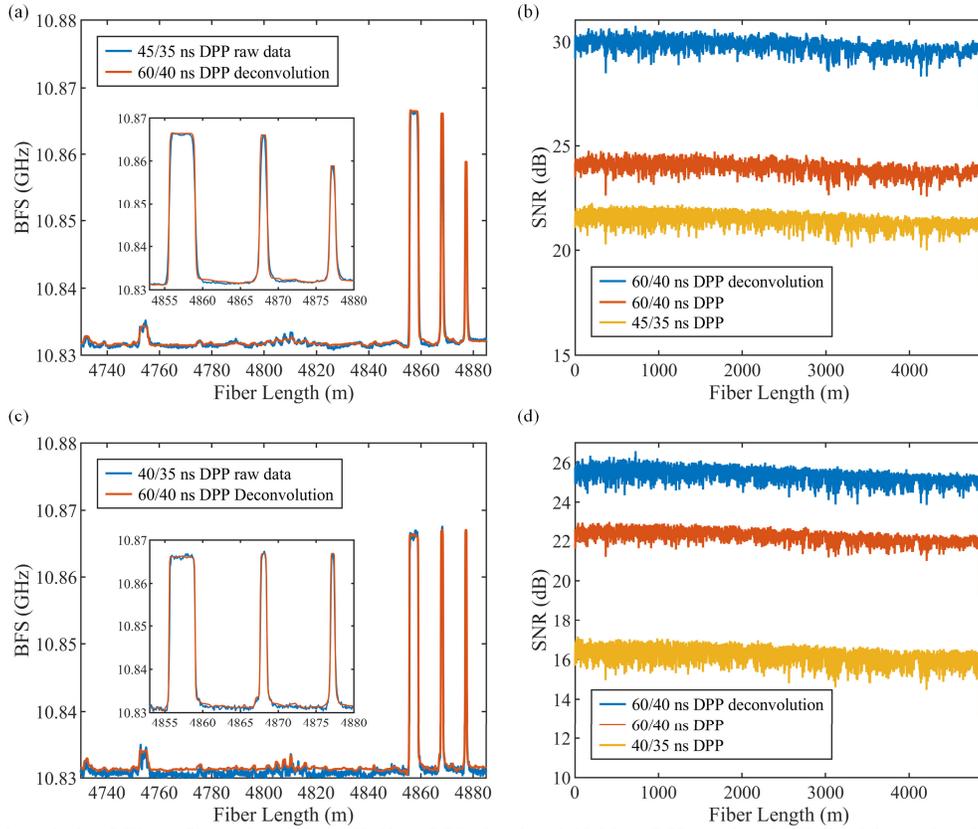

Fig. 9. (a) 1-m spatial resolution BFS profile retrieved from 60/40 ns DPP signal and 45/35 ns DPP results; (b) SNR of the retrieved signal with 1-m spatial resolution, 60/40 ns DPP raw data and 45/35 ns DPP raw data; (c) 0.5-m spatial resolution BFS profile retrieved from 60/40 ns DPP signal and 40/35 ns DPP results; (d) SNR of the retrieved signal with 0.5-m spatial resolution, 60/40 ns DPP raw data and 40/35 ns DPP raw data.

Firstly, deconvolution results with different pump pulse schemes are experimentally validated, which include conventional 20 ns pump pulse and DPP signal with pulse pairs of 60/40 ns, 50/30 ns, 40/20 ns and 30/10 ns, respectively. All the pump schemes have an original 2-m spatial resolution. And the temporal envelope at peak frequency of BGS is used for the deconvolution. The retrieved BFS profiles around the hotspots are shown in Fig. 8. All the hotspots are heated inside the same oven, so they should have the same BFS change. It can be observed that, deconvolution of the 20 ns pump pulse signal will lead to severe distortions ahead the hotspots, together with large BFS fluctuations on the hotspots, especially for the 3-m hotspot. The distorted BFS profile is in good agreement with the simulation, indicating that the distortion is due to the frequency dependency of temporal envelope of Brillouin gain. For the deconvolution results of DPP signal, BFS distortion at hotspots can still be observed as in the 3-m hotspot if the used narrow pulse of the pulse pair is less than 40 ns, e.g. 50/30 ns, 40/20 ns and 30/10 ns. However, if the pulse pair of 60/40 ns is used, the BFS of three hotspots can be distinguished without distortion and presents the best measurement accuracy. This result agrees well with the theoretical prediction.

To further demonstrate the noise reduction capability of the proposed scheme, 1-m spatial resolution BFS profile is retrieved from the 60/40 ns DPP signal. And a 45/35 ns DPP raw data are also used for comparison. As shown in Fig. 9(a), the recovered BFS profile is in good agreement with the conventional DPP result. The inset in Fig. 9(a) shows the BFS profile of the hotspots. The BFS of the 0.5-m hotspot is not fully recovered due to insufficient spatial resolution. The obtained SNR along the fiber is shown in Fig. 9(b). The SNR of the retrieved signal is about 5.9 dB higher than that of the 60/40 ns DPP raw data. Under the same spatial resolution, by using deconvolution algorithm, an 8.4 dB SNR improvement is achieved compared with the conventional 45/35 ns DPP technique. This result indicates that the deconvolution algorithm can increase the spatial resolution meanwhile improve the SNR.

Flexible adjustment of spatial resolution can be achieved by adjusting the regularization parameter $\mu$. Figure 9(c) shows the obtained 0.5-m spatial resolution results from a 60/40 ns DPP signal, together with a 40/35 ns DPP raw data. The BFS of the third hotspot is recovered in the deconvolution profile and agrees well with the 40/35 ns DPP raw data. The small BFS difference at some positions could be attributed to the measurement error of the 40/35 ns DPP signal. The calculated SNR is shown in Fig. 9(d). The signal after deconvolution achieves a 3.2 dB SNR improvement compared with the 60/40 ns DPP raw data, and 9.2 dB SNR improvement over the 40/35 ns DPP raw data.

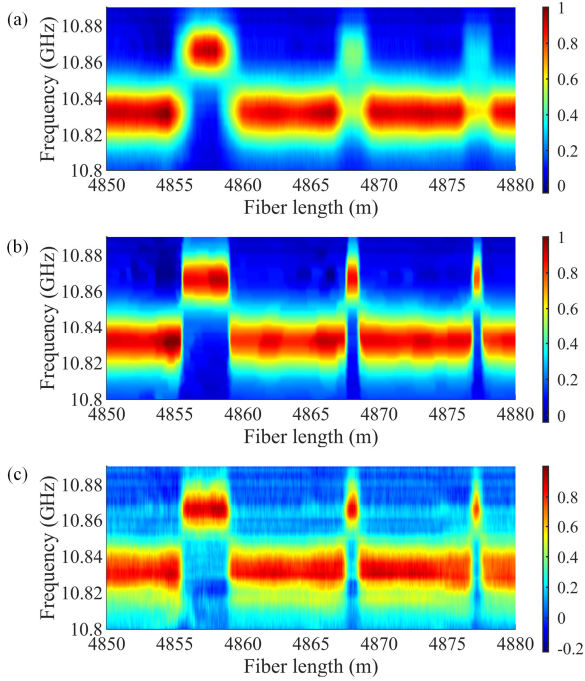

Fig. 10. (a) BGS of the 60/40 ns DPP raw data; (b) BGS of the recovered results with 0.5-m spatial resolution; (c) BGS of the 40/35 ns DPP raw data.

Figure 10 shows the distribution of BGS of the deconvolution and DPP results. The 60/40 ns DPP raw data is shown in Fig. 10(a) and its deconvolution result with 0.5-m spatial resolution is shown in Fig. 10(b). The retrieved BGS exhibits a Lorentzian shape with FWHM of about 27.9 MHz, and three hotspots with sharp edges are recovered. Figure 10(c) shows the BGS from 40/35 ns DPP technique, where worse measurement performance is clearly observed due to system instability. As can be seen, the proposed deconvolution algorithm offers a better robustness compared with the DPP technique, especially at high spatial resolution.

The impact of sampling rate on the recovered results is also experimentally investigated. Figure 11 shows the recovered results of the 0.5 m hotspot with different sampling rates. The regularization parameter $\mu$ is adjusted to achieve identical SNR improvement of 3 dB for the recovered results. The BFS profile with the best accuracy is recovered with 5 GSa/s. And the result indicates that the BFS profiles retrieved from the measurements with 2 GSa/s and 1 GSa/s sampling rate match well with that of 5 GSa/s sampling rate, while the BFS profile retrieved from the measurement with 500 MSa/s turns out to have larger error. The experimental result demonstrates that retrieved accuracy can be improved with higher sampling rate, which is in good agreement with the simulation.

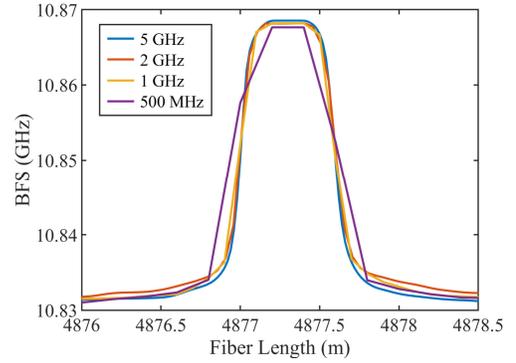

Fig. 11. Retrieved BFS profile for a 0.5-m hotspot with different sampling rates.

## IV. CONCLUSION

In this work, we thoroughly investigate a deconvolution based post-processing method for flexible spatial resolution improvement of BOTDA systems. The frequency dependency of the Brillouin gain temporal envelope and its influence on BFS recovery through deconvolution algorithm is theoretically and experimentally studied. By using the DPP technique with pulse width larger than 40 ns, the BFS distortion around the hotspot after deconvolution is effectively eliminated. To enhance the system robustness against noise, the total variation deconvolution algorithm is introduced. In the experiments, 0.5 m and 1 m spatial resolution is successfully retrieved from a 60/40 ns DPP signal. The recovered signal has a higher SNR and the obtained BFS profile agrees well with the conventional DPP results under identical spatial resolution.

For the conventional DPP technique, the spatial resolution is determined by the pulse pair width difference. In the proposed method, the purpose of DPP pre-processing is to subtract the frequency dependent fluctuation of Brillouin gain. Thus, narrow pulse width difference is not required, which can prevent the related system instability, such as imperfect compensation of polarization fading. Compared with the DPP technique, another advantage of the proposed method is the flexible adjustment of spatial resolution and SNR, thus it can realize adaptive measurement according to the application requirement without reconfiguration of system setup. Moreover, the proposed method can achieve higher SNR over the conventional DPP technique with minor BFS degradation at hotspot. By using higher sampling rate, the BFS degradation can be further reduced to obtain a more precise result. The proposed method is believed to have great potential in high resolution distributed sensing fields with high flexibility and accuracy.

**Li Shen** received the B.S. degree from the School of Optical and Electronic Information, Huazhong University of Science and Technology (HUST), Wuhan, China, in 2016. He is currently working toward the Ph.D. degree with the School of Optical and Electronic Information, HUST. His research interests are optical fiber sensing and optical fiber devices.

**Zhiyong Zhao** received the B.Eng. and Ph.D. degrees from Huazhong University of Science and Technology, Wuhan, China, in 2012 and 2017, respectively. He was a joint Ph.D. Student with the École polytechnique fédérale de Lausanne, Lausanne, Switzerland, from October 2014 to October 2015. He also stayed half a year as a Research Assistant with the School of Electrical and Electronic Engineering, Nanyang Technological University, Singapore. Since June 2017, he has been a Postdoctoral Fellow with the Department of Electronic and Information Engineering, The Hong Kong Polytechnic University, Hong Kong. His current research interests include optical fiber sensing, optical fiber devices, special optical fibers, and nonlinear fiber optics.

**Can Zhao** received the B.S. and Ph.D. degrees from the School of Optical and Electronic Information, Huazhong University of Science and Technology (HUST), Wuhan, China, in 2014 and 2019, respectively. He has been working as a postdoctoral researcher at HUST since 2019. His current research interests include distributed optical fiber sensing and specialty optical fiber.

**Hao Wu** received the B.S., M. Eng., and Ph. D. degrees from the School of Optical and Electronic Information, Huazhong University of Science and Technology (HUST), Wuhan, China, in 2013, 2016 and 2019, respectively. He has been working as a postdoctoral researcher at HUST since 2019. His current research interests include the application of specialty optical fiber and machine learning algorithm in distributed optical fiber sensing.

**Chao Lu** received the B.Eng. degree in electronic engineering from Tsinghua University, Beijing, China in 1985, and the M.Sc. and Ph.D. degrees from the University of Manchester, Manchester, U.K. in 1987 and 1990, respectively. In 1991, he joined the School of Electrical and Electronic Engineering, Nanyang Technological University, Singapore where he was a Lecturer, Senior Lecturer, and Associate Professor until 2006. From June 2002 to December 2005, he was seconded to the Institute for Infocomm Research, Agency for Science, Technology and Research, Singapore, as the Program Director and the Department Manager leading a research group in the area of optical communication and fiber devices. In 2006, he was a Professor with the Department of Electronic and Information Engineering, The Hong Kong Polytechnic University, where he is currently a Chair Professor in Fiber Optics. Over the years, he has published more than 300 papers in major international journals. He has been an organizer or technical program committee member of many international conferences. His current research interests are in the area of high capacity transmission techniques for long haul and short reach optical systems and optical sensing systems.

Prof. Lu is a Fellow of the Optical Society of America.

**Ming Tang** (SM'11) received the B.E. degree from Huazhong University of Science and Technology (HUST), Wuhan, China, in 2001, and the Ph.D. degree from Nanyang Technological University, Singapore, in 2005. His Postdoctoral Research in the Network Technology Research Centre (NTRC) was focused on optical fiber amplifiers, high-power fiber lasers, nonlinear fiber optics, and all-optical signal processing. From February 2009, he was with the Tera-photonics group led by Prof. Hiromasa Ito in RIKEN, Sendai, Japan, as a Research Scientist conducting research on terahertz-wave generation, detection, and application using nonlinear optical technologies. Since March 2011, he has been a Professor with the School of Optical and Electronic Information, Wuhan National Laboratory for Optoelectronics, HUST, Wuhan, China. His current research interests are concerned with optical fiber based linear and nonlinear effects for communication and sensing applications. He has been a member of the IEEE Photonics Society since 2001.